# Observation of robust one-dimensional edge channels in a three-dimensional quantum spin Hall insulator

Shuikang Yu[1,2,#], Junze Deng[3,4,#], Wenhao Liu[5,#], Yunmei Zhang[1], Yiming Sun[1], Nikhil Dhale[5], Sheng Li[5], Wanru Ma[1], Zhuying Wang[1], Ping Wu[1], Zuowei Liang[1], Xuecheng Zhang[1,2], Bing Lv[5*], Zhijun Wang[3,4*], Zhenyu Wang[1,2*] and Xianhui Chen[1,2,6*]

[1]Department of Physics and CAS Key laboratory of Strongly-coupled Quantum Matter Physics, University of Science and Technology of China, Hefei, Anhui 230026, China
[2]Hefei National Laboratory, University of Science and Technology of China, Hefei 230088, China
[3]Beijing National Laboratory for Condensed Matter Physics, and Institute of Physics, Chinese Academy of Sciences, Beijing 100190, China
[4]University of Chinese Academy of Sciences, Beijing 100049, China
[5]Department of Physics, The University of Texas at Dallas, Richardson, Texas 75080, USA
[6]Collaborative Innovation Center of Advanced Microstructures, Nanjing 210093, China

*Correspondence and requests for materials should be addressed to Zhenyu Wang (zywang2@ustc.edu.cn), Zhijun Wang (wzj@iphy.ac.cn), Bing Lv (blv@utdallas.edu) or Xianhui Chen (chenxh@ustc.edu.cn).

**Abstract:**

**Topologically protected edge channels show prospects for quantum devices. They have been found experimentally in two-dimensional (2D) quantum spin Hall insulators (QSHIs), weak topological insulators and higher-order topological insulators (HOTIs), but the number of materials realizing these topologies is still quite limited. Here, we provide evidence for topological edge states within a novel topology named three-dimensional (3D) QSHIs. Its topology originates solely from a nonzero $S_z$ spin Chern number for each $k_z$ plane of the crystal and is realized in bulk α-$Bi_4I_4$ with trivial symmetry indicators, as we show by density functional theory calculations. We experimentally observe the related edge states at each type of monolayer and bilayer step of this material by scanning tunneling microscopy. Consistently, the edge states are neither interrupted, nor backscattered by defects at the step edges corroborating their helical character as expected from the nontrivial topology. Furthermore, two individual edge channels are directly observed at bilayer steps without visible interaction gap opening, demonstrating the robustness of these edge modes against vertical stacking. Our results establish α-$Bi_4I_4$ as the first material realization of a 3D QSHI whose definition goes beyond the scope of topological symmetry indicators, and provide a pathway for realizing nearly-quantized spin Hall conductivity per unit cell in a bulk crystal.**

The QSHI, also known as a 2D $Z_2$-type topological insulator (TI), is insulating in its 2D bulk but harbors conductive edge modes protected by time-reversal symmetry [1-5]. The helical spin-momentum locking of the edge states provides a promising route for realizing highly directional, dense spin currents for information processing and Majorana-excitation engineering for topological quantum computation [6-8]. Therefore, there has been continuous effort to search for QSH states in 2D material systems over the past decade, but thus far, only a handful of candidates, including HgTe quantum wells [9,10], bismuthine [11], stanine [12] and monolayer transition metal dichalcogenides [13-16], have been experimentally verified.

Owing to the complexity and uncertainty in large-scale atomic-thin-layer fabrication, it is desirable to realize robust 1D helical modes at steps in a quasi-3D crystal, ideally irrespective of the number of layers or step geometries. However, this has proven to be difficult: the stacking of QSHI layers could lead to various topological phases of matter, including topological crystalline insulators (TCIs), Weyl semimetals and HOTIs [17-19]. Reducing the interlayer coupling ends up with the idea of weak TIs [20-23], in which the side surfaces are topologically nontrivial with accumulating gapless helical edge states (Fig. 1a), but its material realization is very rare [24-26]. On the spectroscopic side, previous studies have mainly been focused on edge channels at the single-layer limit [11-15, 19, 24, 28-33; summarized in Supplementary Table 1]; the individual edge channels at a multilayer boundary, however, have not been directly visualized, leaving their spatial distribution and interlayer interactions largely unexplored.

Here we use an alternative route to achieve robust helical edge channels in a 3D crystal, that is, *via* a 3D QSHI. The concept of 3D QSHI has been recently established [34, 35] and is defined by a nonzero $S_z$ spin Chern number (SCN) for all the $k_z$ planes, resulting in a non-quantized (but generically nonvanishing) $S_z$ spin Hall conductivity per bulk unit cell; it could be generated by the stacking of QSHIs in the $z$ direction. As a result, some HOTIs (with inversion $Z_4 = 2$) can be classified into the 3D QSHI phase [35]. Taking a step further, the 3D QSHI phase could also be realized in systems with trivial symmetry indicators (36-45), thus termed as a non-symmetry-indicated 3D QSHI. Importantly, when the interlayer interaction is weak and the spin $S_z$ is nearly conserved, the spin $S_z$ Hall conductivity is expected to be nearly quantized with the edge states being slightly gapped out in the 3D QSHI [35]. Therefore, it is critical to check the robustness of helical edge channels at multilayer boundaries experimentally, which could serve as the hallmark of a 3D QSHI.

The material system we focus on is $\alpha$-Bi$_4$I$_4$ bulk crystal which has two Bi$_4$I$_4$ layers in one unit cell. This family of bismuth halides, Bi$_4$X$_4$ (X stands for I or Br), is emerging as a superior platform for designing novel topological quantum phases [32, 33, 46-59]. Monolayer Bi$_4$X$_4$ has been demonstrated to be a QSHI [33, 47, 48], and its stacking generates bulk crystal of $\alpha$ or $\beta$ phase with distinct topological properties [46] (a detailed schematic can be found in Supplementary Note 1). The $\alpha$ phase of Bi$_4$Br$_4$ has recently been confirmed to be a HOTI with $Z_4 = 2$ [32,49-51]. For Bi$_4$I$_4$, while the $\beta$ phase is believed to be a weak TI [52, 53], the α phase is naturally stabilized at temperatures below 300 K by a first-order structural transition. There is an ongoing debate over the topological nature of $\alpha$-Bi$_4$I$_4$: while calculations based on symmetry indicators categorize it as a trivial insulator [37-39,48,52], recent angle-resolved photoemission spectroscopy (ARPES) data and calculations have led to a controversial conclusion suggesting that it is an extrinsic HOTI [53,54]. Using high-resolution STM, we successfully imaged the helical edge channels at all types of monolayer and bilayer steps within a bulk unit cell and characterized the associated density of states (DOS), spatial locations and interlayer interactions. The observation of two individual edge channels on bilayer steps with unchanged DOS inside the bulk gap, together with the nonzero SCN in calculations of the bulk crystal, nail down that $\alpha$-Bi$_4$I$_4$ is in fact a non-symmetry-indicated 3D QSHI. The large bulk band gap around the $E_F$, linear-dispersing edge modes and weak interlayer coupling make $\alpha$-Bi$_4$I$_4$ a promising material system for realizing nearly-quantized spin Hall conductivity per unit cell in a 3D crystal.

We start by checking the structural phase of our cleaved crystals. The building blocks of both the α and β phases are identical, being a layer of weakly-coupled Bi$_4$I$_4$ chains in the *ab* plane (*a* and *b* are conventional lattice vectors with primitive lattice vectors $(a \pm b)/2$). The difference arises from the interlayer stacking: starting from the β phase with one Bi$_4$I$_4$ layer per unit cell (Fig. 1b), in the α phase, the same Bi$_4$I$_4$ layer undergoes an additional shift of $b/2$ every two layers ($t = 2c' + b/2$, where $c'$ denotes the $\beta$-Bi$_4$I$_4$ layer stacking vector) following a BBB'B'-type stacking sequence with a $4c'$ periodicity (Fig. 1c) [54,60]. This stacking sequence results in two distinct cleaved terminations, A and A'. Consequently, a new lattice vector $c$ can be redefined by $c = 2c' + a/2$ in the α phase so that each unit cell contains two layers (B and B') that are reflected to each other under inversion (dashed parallelogram in Fig. 1c). Such a lattice shift, if it exists, can be directly imaged by atomic-resolution topographies across step edges on the (001) surface. Figure 1d depicts a typical atom-registered topography of the cleaved (001) surface; the darker atoms (marked by orange circles) are identified as peripheral bismuth atoms, each of which is covalently bonded to four iodine atoms (blue circles). Interestingly, we do find this additional shift of approximately $b/2$ when crossing one of every two adjacent monolayer steps and across bilayer steps (data and analysis are shown in Supplementary Note 2), which is in good agreement with the stacking sequence of the α phase.

The identification of the α phase is further evidenced by the distinct categories of spectroscopic characteristics of the two different terminations: termination A, which has an equivalent layer beneath, and termination A' which has a shifted layer underneath (Fig. 1c). Figure 1e shows a monolayer step with an expected height of 1 nm, and the differential conductance (d$I$/d$V$), which measures the local DOS, reveals substantial gap features on both sides of the step (Figure 1f; taken far from the step edge). Nevertheless, there is a notable disparity between the gap sizes: surface A has a large insulating gap of 130 meV spanning the Fermi level, whereas a smaller gap of 70 meV is found on surface A'. We emphasize that these two gaps are not due to different set-point conditions of tunneling junctions but have been alternately observed across all monolayer steps. To gain more insight into the electronic structure, we record the spatial variations in the local DOS as a function of energy along a line perpendicular to one monolayer step (Fig. 1h), where electronic standing-wave patterns can be clearly observed near the step edge. The Fourier transform of the local DOS, as shown in Fig. 1i, reveals two main wavevectors that can be attributed to electronic scattering of the conductance band and valence band along the $\Gamma C$ direction respectively (the surface Brillion zone is shown in Fig. 1g). Again, the gap value, defined by the energy spacing between the bottom of the conductance band and the top of the valence band, changes when crossing the monolayer step, confirming the existence of two types of terminations. These two gap values and the band dispersion obtained here are consistent with previous APRES data [53], and a direct comparison is displayed in Supplementary Note 3. We note that while there is a finite DOS at the Fermi level on termination A', the bulk resistivity data of our $\alpha$-Bi$_4$I$_4$ crystals exhibit an insulating behavior (Supplementary Fig. S5). Therefore, we speculate that the finite DOS at the $E_F$ of termination A' is a surface effect due to the coupling to a different underlying layer, which has been commonly observed in other QSH systems [13,14,16, 33].

Having identified the 'bulk' state of one Bi$_4$I$_4$ layer, we next check its bulk-boundary correspondence by imaging the quasiparticle excitations at the step edges. A single layer of Bi$_4$I$_4$ chains on the surface, albeit modified by coupling to the underlying crystal, could be regarded as a 2D QSH system with helical edge states at its boundary. Figures 2a-c show the topography of a monolayer step of termination A and the corresponding edge state in the d$I$/d$V$ maps, respectively. The data are representative of the sample, and set out several key features. First, the spectra away from the step are rather homogeneous, showing bulk energy gaps; additional quasiparticles emerge inside the bulk gap at the edge, and their spectral weight gradually disappears at energies beyond the gap. Second, the DOS associated with the edge state is nearly constant, revealing the fingerprint of a linear Dirac dispersing state in one dimension (Fig. 2d). Third, the DOS variation of the edge channel is extremely small along the step, as expected for suppressed single-particle backscattering (its response to defects is shown in Supplementary Note 4), but perpendicularly confined within a few atomic chains near the step. The characteristic decay length to the bulk state, extracted by an exponential fit, is approximately 1.48 nm (Fig. 2e), similar to that reported in monolayer films [33]. To establish a benchmark for comparison, we calculate the electronic structure of a free-standing Bi$_4$I$_4$ monolayer, and the results are displayed in Fig. 3a. Here, the gap along $\Gamma C$ has been projected to the $\Gamma$ point in the 1D edge Brillouin zone (inset). Despite a slightly larger insulating gap, probably owing to the absence of underneath-layer coupling, the calculation well reproduces the linearly dispersing edge state with a Dirac point located within the gap and the featured flat DOS inside the gap. Remarkably, our calculations further reveal that the wavefunctions of the two edge states dominate at different atomic planes within one Bi$_4$I$_4$ layer on the opposite side (Fig. 3b), which we will discuss in details later.

One unique aspect of this material is the nonvertical facing angle of the step with respect to the (001) surface, so that precisely determining the location of the edge modes is allowed by high-resolution spectroscopic mapping. Since the slope of a step measured via STM is inevitably affected by the tip angle, we determine the step geometry by checking the relative positions of adjacent bismuth and iodine atoms on the (001) surface (for details see Supplementary Note 2), and the result is shown as an inset for each image throughout this work. Two monolayer steps with different structural geometries are shown in Fig. 3c and f, with their corresponding edge modes in Fig. 3d and g. By carefully aligning the atom-registered topographic line profile with the simultaneously acquired d$I$/d$V$ signals, we find that the maximum intensity of the edge channel indeed lies at distinct positions: the edge channel appears at the upper-left corner of the Bi$_4$I$_4$ layer on the steep (left) side, on the right side with a gentle slope it sinks to the bottom corner (Fig. 3e and 3h). These observations are consistent with our theoretical calculations (Fig. 3b). The spectral weight for the edge mode at the bottom corner is weaker, probably due to a smaller tunneling possibility to the STM tip. We note that the edge modes have been checked at all types of monolayer steps for both terminations A and A' with different step geometries, and they all obey this spatial distribution rule (Supplementary Note 5). The unique locations of these edge modes, as predicted by calculations, manifest their topologically nontrivial origin.

Having characterized the edge states at monolayer steps, we turn to a central issue in α-Bi$_4$I$_4$: what happens at the bilayer steps? From the perspectives of the Z$_2$ index and symmetry indicators, an even-layer system could be topologically trivial, and interlayer coupling can obliterate the helical edge states. Specifically, high-throughput screening calculations suggest that bulk α-Bi$_4$I$_4$ should be classified as a trivial insulator with vanishing symmetry indicators $(z_{2w,1}, z_{2w,2}, z_{2w,3}, z_4) = (0,0,0,0)$ and zero mirror Chern numbers [37-39,61], which has been crosschecked with our DFT calculations. However, our STM experiments reveal robust edge modes at all types of bilayer steps (BB'-type and BB-type), calling into question this conviction. Figure 4a-f shows the edge states of BB'-type bilayers (termination A') on both the left and right sides, and additional data sets can be found in Supplementary Note 6. By analyzing the spatial distribution of the edge states, we indeed find two well-separated DOS maxima, instead of one, at a bilayer step. The two channels on the right side (with gentle slope) are situated halfway of the step and at the bottom corner, respectively, whereas on the left (steep) side, the two are roughly located at the top corner and middle of the step. Because the lower channel on the steep side is buried under the surface, as a surface probe STM measurements might not reveal its precise position. However, the observation of two edge channels is robust in our data (Supplementary Note 6), and the spacing between them is approximately half of the width of the step. Moreover, the spectral weight of the edge states on the left side is stronger, similar to that of a monolayer step. These observations are consistent with each other and strongly indicate that a bilayer step can be regarded as two monolayers being simply stacked together.

One might doubt that the double edge channels at a bilayer step could arise from the multiple-tip effect. However, we can exclude this possibility by the following observations. First, before taking dI/dV measurements, we have carefully checked the topographic scans of these steps to confirm that they are atomically sharp (Supplementary Fig. S9; it is worth noting that the topographic line prolines in Figs. 3 and 4 are extracted from dI/dV maps, so the step edges appear broader than those in topographic scans). Second, dI/dV maps taken in a field of view including both monolayer and bilayer steps are shown in Fig. S10. While two edge channels can be clearly observed at the bilayer step, there is only one channel at monolayer steps with the same scanning tip. Third, although the broadening of a step in topographic images is inevitably affected by the shape of the tip and the feedback circuit settings, in the simultaneously acquired d$I$/d$V$ maps, the edge channels are always located at these characteristic positions as mentioned above, and their distance is approximately half of the width of the step, which cannot be a coincidence.

To further characterize the inter-edge coupling at a bilayer step, we display the normalized tunneling spectra taken at several bilayer edges in comparison to those at monolayer steps in Fig. 4g. If relevant, the interlayer coupling will lead to an interaction gap in the 1D edge modes. However, we do not find any notable difference in their spectral line-shapes taken at the monolayer and bilayer steps, indicating a rather weak coupling of the edge states along the stacking direction. We note that sometimes a tiny dip appears at approximately -0.2 eV in the spectra, but it appears even at the monolayer steps (Fig. S8), which is unlikely to be an interaction gap. Moreover, this feature is buried in the bulk valence band and therefore has little impact on the application of topological edge states. Based on the experimental energy resolution, the gap induced by interlayer coupling, if exists, should be smaller than 4 meV (Methods). This is basically consistent with our first-principles calculations, which show a very tiny gap (~ 6 meV) opening on the (100) side surface (Fig. 4h-j) due to interlayer coupling. In contrast, the in-plane hybridization can completely gap out the edge states on a narrow terrace with a width of approximately 2-3 nm as shown in Fig. S7. Therefore, the remaining lateral distances (approximately 2nm; Fig. S9) between the pairs of edge states at bilayer steps can be excluded as an origin of the missing gap in the edge state. Consequently, α-Bi4I4 can be viewed as a bulk stack of 2D QSH systems with helical edge channels encircling each layer.

The observation of robust helical edge modes calls for a new topological category for this material beyond symmetry-indicator classification theory [37-45] or topological quantum chemistry [36]. From our first-principles calculations, we conclude that $\alpha$-Bi4I4 is a 3D QSHI with trivial (zero-valued) symmetry indicators, but is characterized by nonzero SCNs [34,35]. The SCN is defined on the $S_z^\pm$ branch of occupied states, with $S_z^\pm$ the eigenvalues of the spin $S_z$ projector $[S_z(\mathbf{k})]_{mn} \equiv \langle u_m(\mathbf{k})|S_z|u_n(\mathbf{k})\rangle$, where $m$ and $n$ run over the occupied states. The SCN can be computed by the spin-resolved $k_a$-directed Wilson loop spectrum for the $S_z^+$ sector. The $S_z^\pm$ SCNs are obtained to be $\pm 2$ in the spin-resolved Wilson loop spectrum for all the $k_z$ planes, and the results for the $k_c = 0$ and $k_c = \pi$ planes are presented in Fig. 4k and l, respectively, indicating the presence of nonzero SCNs ($C_s^\pm = \pm 2$). Therefore, $\alpha$-Bi4I4 can be classified as a non-symmetry-indicated 3D QSHI, but not a helical HOTI [35]. The near quantization of spin Hall conductivity can be achieved depending on the gap size of the edge states and the strength of the interlayer coupling.

In fact, our results provide key information on the weak interlayer coupling in $\alpha$-Bi4I4. First, in both tunneling spectra and side-surface-state calculation, the gap opening on the 1D edge state due to interlayer coupling should be very tiny (Fig. 4g and i). Second, the thickness of a single Bi4I4 layer is approximately 1 nm, and the edge states are confined on the lower-right/upper-left corner (as plotted in Fig. 3b), which is further verified experimentally (Figs. 3c-h and 4a-f). Since the edge states of the 'thick' Bi4I4 monolayer are not at the same $z$ value, the $z$-directed electric field can easily split these edge states of opposite edges (Supplementary Note 7), which is quite different from the case for atomically thin monolayers such as graphene. Third, the spin textures of the edge states in the bulk energy gap are almost aligned along $\pm S_Z$ ($0.93 \times \hbar/2$) with spin-momentum locking, leading to negligible hybridization between opposite spin channels of different layers. The well-localized edge states in the $z$ direction significantly weaken the interlayer edge state coupling. When the thickness of $\alpha$-Bi$_4$I$_4$ is reduced to few layers, a multilayer QSHI could be achieved that is featured by a layer-dependent high SCNs (62). Interestingly, it has been shown in ref. 62 that the bulk-boundary correspondence is still robust if considering the spin (feature)-spectrum topology, even though the interlayer coupling slightly gaps the edge states.

In summary, the combination of our STM and first-principles calculations establishes α-Bi$_4$I$_4$ bulk crystal as a non-symmetry-indicated 3D QSHI harboring helical edge channels at the boundary of each Bi$_4$I$_4$ layer. Excitingly, spectroscopic features at bilayer steps confirm the robustness of these edge modes against stacking, providing a looser constraint for the realization of a (nearly) quantum spin Hall effect. Our findings also unveil a nontrivial example of 3D QSHIs with vanishing SIs, characterized by nonzero $S_Z$ spin Chern numbers instead.

**Methods**

**Crystal growth:** Single crystals of Bi$_4$I$_4$ were synthesized through a chemical vapor transport reaction. Bismuth chunks (Alfa Aesar, 99.999%) were combined with HgI$_2$ powder (Alfa Aesar, 99%+) at a stoichiometric ratio of 1:1. This mixture was placed in an evacuated quartz tube and then placed into a horizontal tube furnace. The material end of the quartz tube was placed at 270 °C, while the cold end was placed at 210 °C. After reacting for two weeks at this temperature gradient, black needle shaped crystals were obtained. The whole tube without an opening was further annealed at 160 °C for 8-10 weeks to improve the homogeneity, quality, and size of the crystals. The grown crystals have a typical length of 8-10 mm, with a width of 0.5 mm on the (*00l*) surface and a thickness of approximately 0.2 mm on the (*l00*) surface.

**STM experiments.** The STM data were acquired using a commercial CreaTec low temperature system. The single crystals were cleaved *in situ* under cryogenic ultrahigh vacuum at $T \sim 80$ K and then immediately inserted into the STM head that is held at 4.3 K. PtIr tips were used and tested on the surface of single crystal Au (111) prior to the measurements. The majority of spectroscopic data were acquired by the standard lock-in technique at a frequency of 987.5 Hz with a modulation bias of 2.5 meV. This yields an energy resolution of $\Delta E_{mod} = \sqrt{2}V_m \sim 3.54$ meV. Taking into account the thermal broadening $\Delta E_{thermal} \approx 3.53 k_B T \sim 1.3$ meV, the total energy resolution is given by $\Delta E_{total} = \sqrt{\Delta E_{thermal}^2 + \Delta E_{mod}^2} \approx 3.77\ meV$.

**DFT Calculations.** We performed first-principles calculations based on density functional theory using the projector augmented wave (PAW) method [63,64] implemented in the Vienna ab initio simulation package (VASP) [65, 66] to obtain the electronic structures. The generalized gradient approximation (GGA) with the Perdew, Burke and Ernzerhof (PBE) for the exchange-correlation functional [67] was adopted. The kinetic energy cutoff was set to 500 eV for the plane wave bases. The BZ was sampled by the Γ-centered Monkhorst-Pack method [68] with a 9×9×6 k-mesh in the self-consistent process. The irreducible representations of electronic states are obtained by IRVSP [69]. The maximally localized Wannier functions (MLWFs) under the basis of the Bi-p and I-p orbitals are extracted from DFT calculations [70-73]. The surface states are calculated via WannierTools [74].

## Acknowledgments

We sincerely thank Tay-Rong Chang, Kuan-Sen Lin, Barry Bradlyn and Songbo Zhang for the inspiring discussions. Work at USTC is supported by the National Key R&D Program of the MOST of China (grants no. 2022YFA1602600), the National Natural Science Foundation of China (grants nos. 11888101, 12074364 and 52261135638), the Innovation Program for Quantum Science and Technology (grant no. 2021ZD0302802), the Anhui Initiative in Quantum Information Technologies (grant no. AHY160000), and Systematic Fundamental Research Program Leveraging Major Scientific and Technological Infrastructure, Chinese Academy of Sciences under contract No. JZHKYPT-2021-08. J. Deng and Zhijun Wang were supported by the National Natural Science Foundation of China (grants nos. 11974395 and 12188101), the Strategic Priority Research Program of Chinese Academy of Sciences (grant no. XDB33000000), the National Key R&D Program of China (grants nos. 2022YFA1403800 and 2022YFA1403400), and the Center for Materials Genome. Work at University of Texas at Dallas acknowledge the University Research Enhancement Fund.

## Competing financial interests

The authors declare no competing interests.

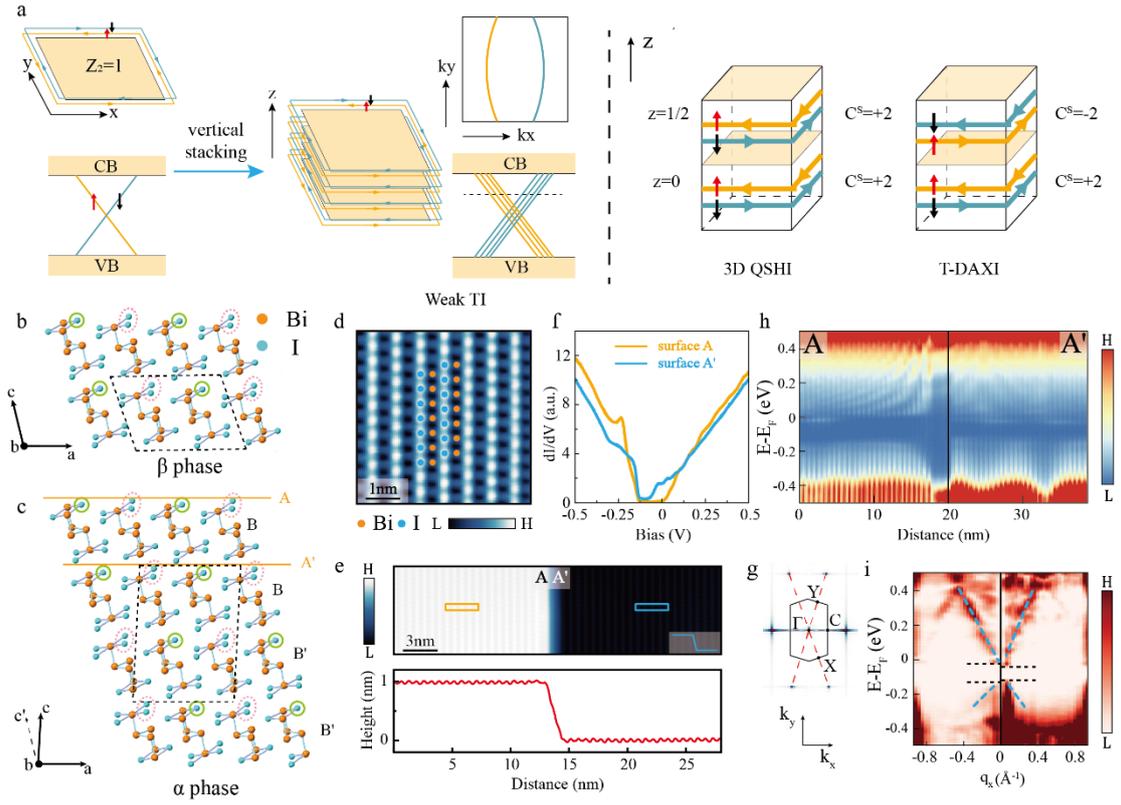

**Figure 1. Concept of 3D QSHI and identification of the structure. a,** The stacking of 2D QSHI layers can lead to a weak TI that has a single 2D TI layer in one unit cell (left panel). A 3D QSHI consists of two QSHI layers in a unit cell that are characterized by the same SCNs, while two QSHI layers with opposite SCNs result in a time-reversal-invariant double axion insulator harboring quantized nontrivial partial axion angles (ref. 35; right panel). **b, c,** Crystal structures of β- and α-$Bi_4I_4$, respectively. Green circles and pink ovals mark the two types of $Bi_4I_4$ chains that have an $b/2$ shift along the $b$ axis; the dark dashed parallelograms denote the crystalline unit cell. **d,** STM topography showing the positions of bismuth (yellow dots) and iodine atoms (blue dots). **e,** Topography including a monolayer step with a height of 1.04 nm. **f,** dI/dV spectra obtained at terminations A and A'. These spectra are averaged in the regions marked with blue and yellow boxes in **e**. **g,** The 2D Brillouin zone of the (001) surface. **h,** Spatial differential conductance along the linecut across a monolayer step showing the standing wave patterns. **i,** Fourier transform of **h** showing the quasiparticle interference signal along $\Gamma C$. The blue dashed lines roughly track the scattering signal, and the black dashed lines mark the bulk energy gaps. STM setups: sample bias $V_s$ = 500 mV; d, tunneling current $I_t$ = 200 pA; e, $I_t$ = 100 pA; f, $I_t$ = 300 pA, modulation bias $V_m$ = 2.5 mV; h, $I_t$=600 pA, $V_m$=5 mV.

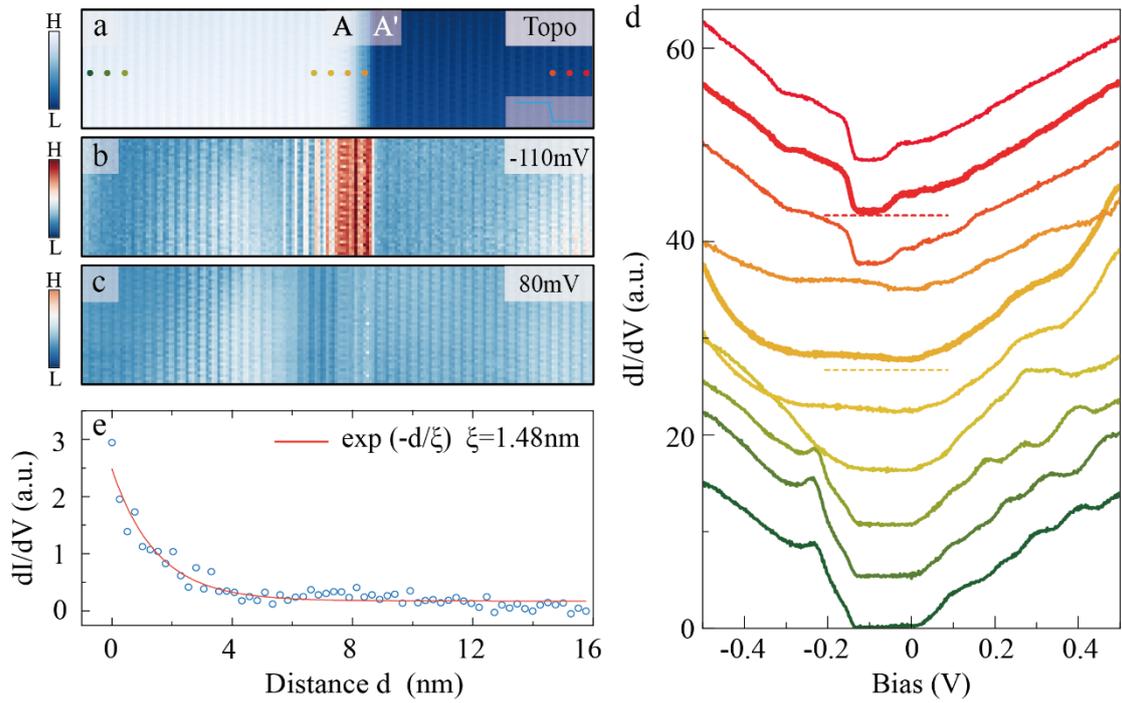

**Figure 2. Helical edge states at the monolayer step. a,** Topographic image of a monolayer step. Left, termination A; right, termination A'. **b, c,** Corresponding d$I$/d$V$ maps at -110 mV (within the bulk gap) and 80 mV (outside the gap), respectively. **d,** Tunneling spectra taken at the corresponding positions marked with colored dots in **a**. The spectra are offset for clarity and the horizontal dashed lines denote the zero differential conductance level for spectra of the edge state (yellow) and termination A'(red). **e,** Intensity distribution of the differential conductance at -110 meV as a function of the distance from the edge of termination A. The red curve is an exponential fit to the data with a characteristic decay length of 1.48 nm. STM setups: $V_s$ = 500 mV; a, $I_t$ = 400 pA; b, c, $I_t$ = 400 pA, $V_m$ = 5 mV; d, $I_t$ = 300 pA, $V_m$ = 2.5 mV.

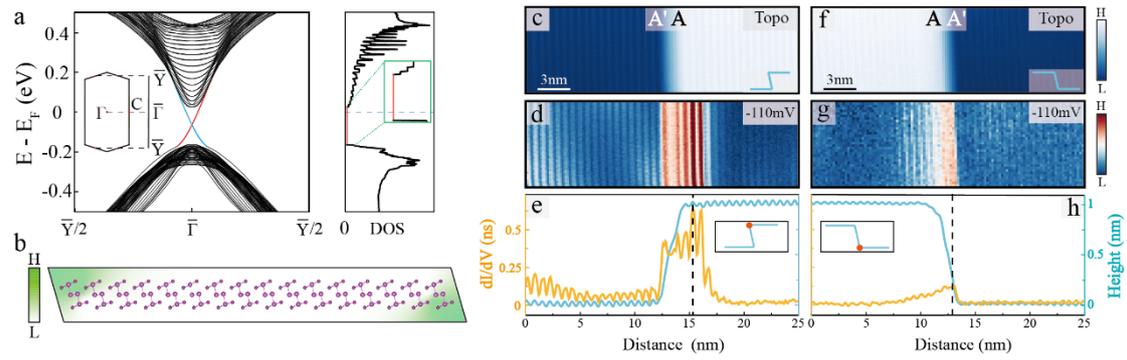

**Figure 3. Spatial location of the edge modes. a**, Calculated band structure for a free-standing $Bi_4I_4$ monolayer. The topological edge state emerges inside the bulk gap and is coded in red/blue with $<\pm S_z>$. The right panel shows the associated electronic DOS. **b**, Calculated distribution of the edge state wavefunction within a $Bi_4I_4$ monolayer: it is confined at the lower-right/upper-left corner of the layer. **c, f,** STM topographies of monolayer steps (termination A). **d, g,** Corresponding differential conductance maps inside the bulk gap (-110 meV) showing the edge modes. **e,** Atom-registered line profiles of simultaneously acquired morphology and differential conductance across the monolayer steps, which directly reveal the positions of the edge modes**.** The dashed lines mark the maxima in the differential conductance. Insets are sketches showing the positions of the edge modes. STM setups: $V_s$ = 500 mV; **c-h**, $I_t$ = 300 pA, $V_m$ = 5 mV;

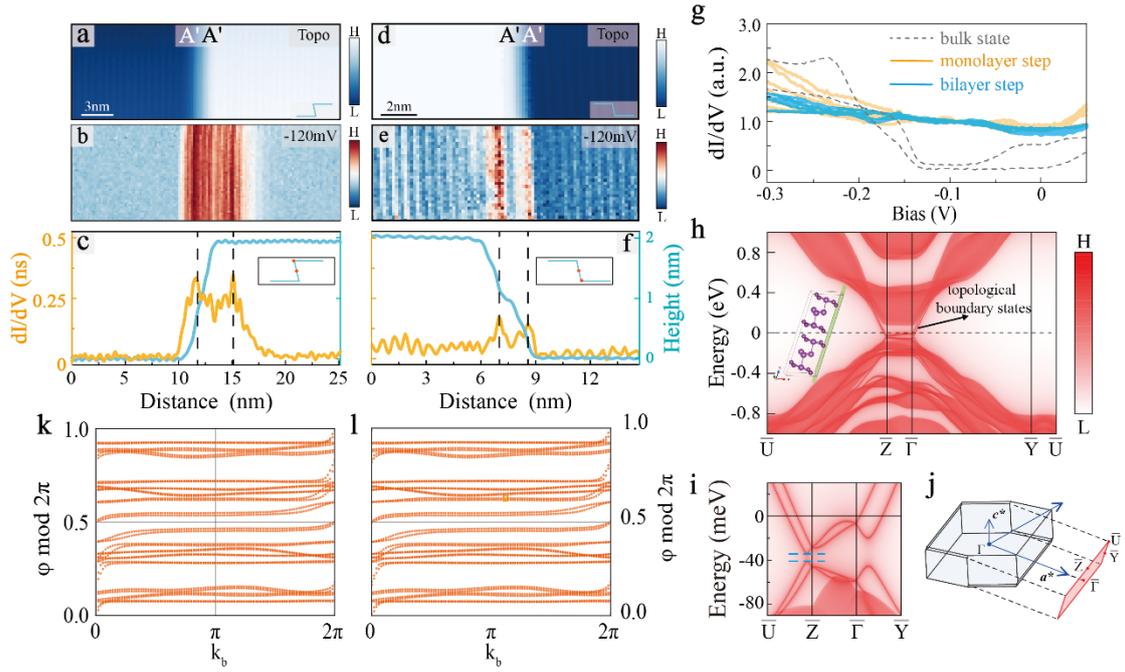

**Figure 4. Robust edge modes at bilayer steps and the topological category of α-Bi$_4$I$_4$. a, d,** STM topographies of BB'-type bilayer steps (both sides are termination A'). **b, e,** Corresponding differential conductance maps showing the robustness of the edge modes where two individual channels can be clearly found. **c, f,** Atom-registered line profiles of simultaneously acquired morphology and differential conductance across the bilayer steps. Two separate maxima can be found near a single step, and their positions are marked by dashed lines. **g,** Direct comparison of the normalized spectra taken at monolayer (yellow) and bilayer (blue) steps, together with the spectra far from step edges (dashed dark line). **h,** Calculated topological boundary state on the (100) side surface, showing a very tiny gap due to interlayer coupling. The insert illustrates the (100) side surface of α-Bi$_4$I$_4$. **i,** Zoomed-in view of the topological boundary state shown in **h**, where the tiny gap is marked by blue dashed lines. **j,** Bulk and projected (100) surface Brillouin zones of α-Bi$_4$I$_4$. **k, l,** SCNs computed by the spin-resolved $k_a$-directed Wilson loop spectrum for the $S_z^+$ sector at the $k_c = 0$ (**k**) and $k_c = \pi$ (**l**) planes, respectively. The results indicate nonzero SCNs $C_s^\pm = \pm 2$ in both planes. STM set: $V_s$=500 mV, **a-c,** $I_t$=250 pA, $V_m$=5 mV; **d-f,** $I_t$=250 pA, $V_m$=5 mV; **g,** $V_m$=2.5 mV.